\documentclass[conference]{IEEEtran}
\IEEEoverridecommandlockouts

\usepackage{cite}
\usepackage{amsmath,amssymb,amsfonts}
\usepackage{algorithmic}
\usepackage{graphicx}
\usepackage{textcomp}
\usepackage{xcolor}
\usepackage{amsmath,amssymb,amsfonts}
\usepackage{algorithmic}
\usepackage{colortbl}
\usepackage{caption}
\usepackage{subcaption}
\usepackage{wrapfig}
\usepackage[utf8]{inputenc}
\usepackage{algorithm}
\usepackage{dblfloatfix}
\usepackage{mathtools}

\usepackage{authblk}

\newcommand*{\affaddr}[1]{#1} 
\newcommand*{\affmark}[1][*]{\textsuperscript{#1}}

\begin{document}
\newcommand{\etal}{\textit{et al.}}
\title{Trust Computational Heuristic for Social Internet of Things: A Machine Learning-based Approach}
\author{%
Subhash Sagar\affmark[1,*]\thanks{\emph{*Corresponding Authors: (subhash.sagar, adnan.mahmood)@mq.edu.au}}, Adnan Mahmood\affmark[1,*], Quan Z. Sheng\affmark[1], and Wei Emma Zhang\affmark[2]\\
\affaddr{\affmark[1]Department of Computing, Macquarie University, Sydney, NSW 2109, Australia }\\
\affaddr{\affmark[2]School of Computer Science, The University of Adelaide, Adelaide, SA 5005, Australia}\\
}




\maketitle

\begin{abstract}
The Internet of Things (IoT) is an evolving network of billions of interconnected physical objects, such as, numerous sensors, smartphones, wearables, and embedded devices. These physical objects, generally referred to as the \emph{smart objects}, when deployed in real-world aggregates useful information from their surrounding environment. As-of-late, this notion of IoT has been extended to incorporate the social networking facets which have led to the promising paradigm of the \emph{`Social Internet of Things' (SIoT)}. In SIoT, the devices operate as an autonomous agent and provide an exchange of information and services discovery in an intelligent manner by establishing social relationships among them with respect to their owners. Trust plays an important role in establishing trustworthy relationships among the physical objects and reduces probable risks in the decision making process. In this paper, a trust computational model is proposed to extract individual trust features in a SIoT environment. Furthermore, a machine learning-based heuristic is used to aggregate all the trust features in order to ascertain an aggregate trust score. Simulation results illustrate that the proposed trust-based model isolates the trustworthy and untrustworthy nodes within the network in an efficient manner.
\end{abstract}


\section{Introduction}

Over the past decade, the notion of Internet of Things (IoT) has evolved as a new generation of a network of billions of devices seamlessly connected over the Internet. These devices are regularly outfitted with sensors which monitor different aspects of human life for supporting numerous beneficial applications and services \cite{ATZORI20102787}. A big value of the IoT resides on its ability to create a network of resources, i.e., by making resources social, where social relationships facilitate the discovery of resources that have the capabilities required to solve a particular task. To accomplish this goal, IoT should be endowed with the ability to define and manage social relationships between resources in every aspect \cite{Mendes2011SocialdrivenIO}. The promising paradigm of Social Internet of Things (SIoT) has transpired recently which can be seen as the combination of social networks and IoT, wherein every object is capable of establishing social relationship autonomously with the other objects depending on the rules set by their respective owners \cite{ATZORI20123594}. Nevertheless, the convergence of physical objects, humans, and cyber components in SIoT presents new concerns for risk, privacy, and security. As the intent of SIoT services is to make decisions autonomously without the need for any human intervention, the notion of \emph{trust} is recognized as a prospective solution for supporting both humans and services in order to overcome the perception of insecurity and minimize the risks when making a decision. 

\begin{figure}[h]
    \centering
    \includegraphics[width=3in, height=2.8in]{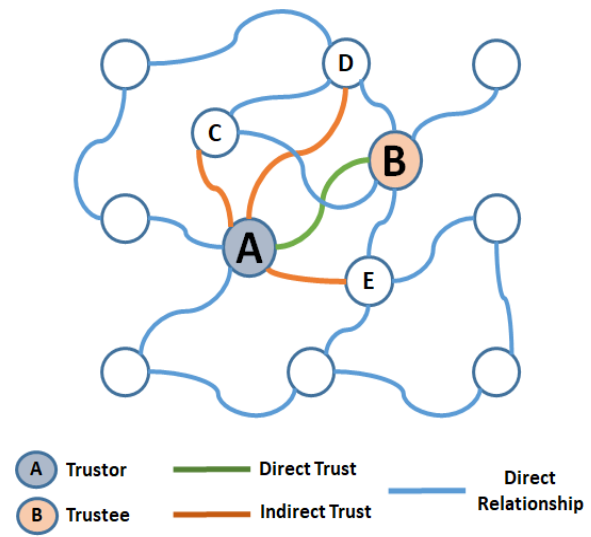}
    \caption{A Schematic Representation of Trust Computation in a SIoT Environment}
    \label{fig:iot_network}
\end{figure}

The paradigm of trust has been used in several disciplines, including but not limited to, psychology, sociology, and computer science \cite{6362662, sociology}. In terms of SIoT, trust can be referred to as a notion of \emph{`belief'}  or \emph{`confidence'} of a trustor in a trustee to perform a specific task for satisfying a trustor's expectations in a specific context within a particular period of time \cite{Amin2019}. As depicted in Fig. \ref{fig:iot_network}, if a node A (trustor) needs to compute the trust value for node B (trustee), it computes the direct trust on its own and requests the mutual friends (e.g., nodes C, D, and E) to provide their recommendations for the same. Finally, the combination of direct and indirect trust provides an aggregate trust score. 


The main reason for providing the trust management system for SIoT is apparent since there are misbehaving nodes which may perform different types of attacks (such as ballot-stuffing attack, bad-mouthing attack, self-promoting and whitewashing attack) based on their social relationships with other nodes for their malicious advantages at the expense of other IoT devices which provide similar SIoT services.

A number of trust management schemes have been proposed in recent years. The authors in \cite{Nitti2014} delineated a subjective and objective trustworthiness scheme in order to ascertain trust in a SIoT environment, wherein trust of any particular node was evaluated by aggregating three salient features, i.e., centrality, opinions from its common friends, and the direct experiences. Nevertheless, the influence of each of these features on the trust aggregation process was ascertained by certain weighting factors that are themselves extremely difficult to figure primarily owing to the reason that trust depends on several complex parameters, i.e., context, time, resources, and the environment. Similarly, the authors in \cite{Chen2016} proposed an adaptive trust protocol for a SIoT system. It employs direct monitoring and indirect observations from users with similar social activity metrices, i.e., social contacts, honesty, and communities-of-interest, to ascertain overall trust. Furthermore, in order to aggregate the same, an adaptive filtering method was designed to integrate direct and indirect observations with weighting parameters for each observation. However, the authors did not validated their envisaged protocol on a wide range of dynamic environmental scenarios, wherein assigning the weighting parameters is itself a complicated chore. In \cite{BenAbderrahim2017}, the authors employed community-of-interest as a social trust metric to envisage a trust management mechanism for purposes of IoT, wherein a Kalman filter was used as a tool to estimate the trust value of a node before interaction. However, the aggregate trust value was calculated using a linear equation with weighting factors for both direct and indirect trust. 

In \cite{Truong2017}, the authors presented a subjective trust model employing the social features of similarity in respect of common interests, honesty, and cooperativeness for evaluating the trust score of a particular node. To obtain the direct trust, a weighted sum metric was used to aggregate both the current as well as past experiences. Nevertheless, the model did not account for the indirect observations or recommendations from other nodes in the network which is essential for the IoT services delivery. Moreover, in \cite{7496623}, a context-based social trust model has been developed for IoT purposes by considering social relationships amongst nodes. For computing the trust score of a node, both direct and indirect observations with a transaction context was used. However, to obtain a single trust score, a weighted sum metric was used for aggregation. Recently, the authors in \cite{8364607} suggested a trust framework model based on the social profile of a node, wherein different social features were accumulated to ascertain the trust score of a node. In addition, a machine learning-based algorithm was exploited to aggregate only the direct trust metric and which is not sufficient enough to decide whether a node is trustworthy or not. 

In this research work, we have computed trust based on both the direct observation as well as the indirect recommendations. Our primary contributions in this paper are threefold: 
\begin{itemize}
  \item A comprehensive trust model for the SIoT environment has been envisaged which specifies the formation of trust via both direct and indirect observation of the nodes; 
  \item A Machine Learning (ML)-based aggregation scheme has been envisaged in contrast to the conventional trust-based heuristics to aggregate the trust attributes for obtaining a single trust score; and
  \item Performance evaluation of the proposed model has been comprehensively carried out in a simulation environment.
\end{itemize}

\section{Trust Computation Model}



\begin{figure}[h]
    \centering
    \includegraphics[width=\linewidth, height=2.05in]{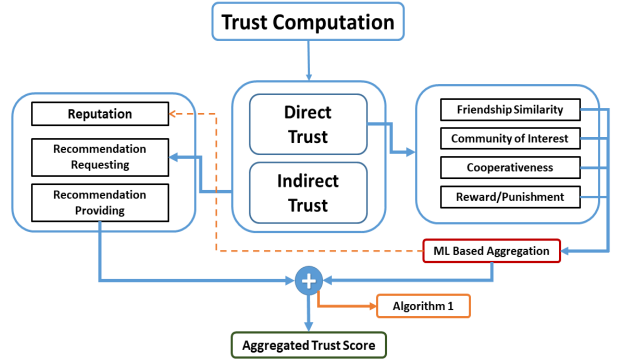}
    \caption{The Proposed Trust Computation Model}
    \label{fig:trust_model}
\end{figure}

This section presents the details of the computational model for an efficient and flexible trust management scheme in a SIoT environment. Our trust model in this paper comprises of two metrics, \textit{Direct Trust Metric (DTM)} and \textit{Indirect Trust Metric (ITM)}, as depicted in Fig. \ref{fig:trust_model}, where DTM gives the notion of direct observation while ITM provides the reputation of nodes in the network. The trust assessment for node $i$ (trustor) towards node $j$ (trustee) is denoted by $T_{X}(i,j)$ where \emph{X} denotes the social attributes like $Friendship \ Similarity, \ Community-of-Interest, \ Reward,$ and $Cooperativeness$. The range of $T_{X}(i,j)$ varies from $[0, 1]$ where values nearer to $\textit{0}$ indicate untrustworthiness while values around $\textit{1}$ indicate trustworthiness. After aggregating all the features $T_{X}(i,j)$ from the direct interaction via ML-based algorithm, the result is stored in the repository and is used as a direct trust score. For indirect trust (recommendations), the node (trustor) requests the direct trust from the other nodes. Subsequently, Algorithm 1 is utilized to combine both the results (trusts) to obtain the final trust score.

\subsection{Direct Trust Metric (DTM)} DTM is used to provide direct observation of a trustee prior to interaction. Although a trustee can be assessed via numerous different attributes, in this paper, we have employed four main attributes for the assessment of any trustee with respect to the trustor and which are elaborated as follows:

\subsubsection{Friendship Similarity (FS)}
Friendship similarity represents social relationship in terms of interaction among participating objects. It measures the importance of an object among other objects with reference to a specific task and context. This property of an object is ascertained as:

\begin{equation}
    T_{FS} (i,j) = \frac{|F_i \cap F_j|}{|F_i| - 1} \label{eq:fs}
\end{equation}
where, $F_i$ and $F_j$ refers to a set of friends of node $i$ and node $j$ respectively and $|.|$ shows the cardinality of a set.  

\subsubsection{Community-of-Interest (CoI)}
This kind of attribute represents the similarity of nodes with respect to the social interest communities or groups. Therefore, nodes with high CoI have more chances of interacting with each other in order to develop a trustworthy relationship. CoI-based trust between two nodes is computed as:

\begin{equation}
    T_{CoI} (i,j) = \frac{|C_i \cap C_j|}{|C_i|}  \label{eq:coi}   
\end{equation}
where, $C_i$ and $C_j$ represents set of communities of node $i$ and node $j$ respectively. 
    
\subsubsection{Cooperativeness (CoP)} CoP manifests whether a trustee is socially cooperative with a trustor or not. Since CoP refers to a measure of balance in the interaction between the nodes, we can employ the entropy function delineated in \cite{5484757} to calculate CoP-based trust as: 

\begin{equation}
    T_{CoP} (i,j) = -T_p \ log (T_p) - (1-T_p) \ log (1-T_p) \label{eq:cop}   
\end{equation}
where, $T_p$ represents fraction of messages during the interaction. 

\subsubsection{Reward/Punishment} In order to maintain both trustworthy relationships and punish misbehaving nodes, we utilize an exponential downgrading formula to provide the incentive to honest nodes and penalties to misbehaving nodes as:

\begin{equation}
    T_{Reward} (i,j) = \frac{|Int -Int_{U}|}{|Int|} \ e^ {-{({\frac{|Int_{U}|}{|Int|}})}} \label{eq:reward}
\end{equation}
here, $Int$ highlights the total number of interactions and $Int_{U}$ refers to the count for the number of unsuccessful interactions between node $i$ and node $j$.

Traditionally, to aggregate the overall trust, a linear equation with the weighting factor is used as shown in Eq. \eqref{eq:total}, however, this approach has numerous disadvantages and challenges (as discussed in Section I) while determining the appropriate value of weights.  

\begin{equation}
\begin{multlined}
T_{Direct} (i,j) = w_1 T_{FS} (i,j) + w_2 T_{CoI} (i,j) + {}\\
                   w_3 T_{CoP} (i,j) + w_4 T_{Reward} (i,j) \label{eq:total}
\end{multlined}
\end{equation}

Therefore, to overcome this drawback, we hereby propose a new machine learning-based approach which combines direct and indirect trust to ascertain an overall trust value. Additionally, this approach also identifies the impact of each of these features on the aggregated trust value. 

\subsection{Indirect Trust Metric (ITM)}

A reputation metric (ITM) is employed in order to ascertain the trustee based on the opinion of other nodes in the network. Nevertheless, the reputation of objects vary from node to node, and therefore, it is not optimal to take account of all the nodes in the network for computing the reputation of a trustee. Thus, in this paper, reputation value is requested from nodes having at least a single friend in common between trustor and trustee.

\begin{algorithm}[h]
\caption{Trust Score Estimation}
\begin{algorithmic}[1]
  \STATE $ 0 \rightarrow Untrustworthy, 1 \rightarrow Trustworthy, 2 \rightarrow Neutral$
  \STATE \textbf{Input}:  Direct Trust \{0, 1 or 2\}, Recommendations \{ $|T|$, \ $|U|$, and $|N|$ \},  
  \STATE \textbf{Output}: Single Trust Score \{0 or 1\},
  \STATE $|T| \rightarrow No: \ of \ Trustworthy \ Recommendations$,  
  \STATE $|U| \rightarrow No: \ of \ Untrustworthy \ Recommendations$, 
  \STATE $|N| \rightarrow No: \ of \ Neutral \ Recommendations$,
  \STATE $\theta \rightarrow \ Threshold \ (\%)$,  
  \STATE $P_U \rightarrow Untrustworthy \ Recommendations \ (\%)$,
  \STATE $P_T \rightarrow Trustworthy \ Recommendations \ (\%)$ 
  \IF {There are no Recommendations}
        \STATE $Final \ Trust = Direct \ Trust$ 
  \ENDIF
  \IF {$Direct\ Trust == 0$} 
        \IF {$ (|U| >= |T| \ \vert \vert \ (|N| >= |T| \ \&\& \ |N| >= |U|)$}
            \STATE Node is Untrustworthy
        \ELSE
            \STATE $P_T = \frac{|T|}{Total \ Recommendations + 1}$
            \IF{$P_T > = \theta$}
                \STATE $Node \ is \ \textbf{Trustworthy}$
            \ELSE
                \STATE $Node \ is \ \textbf{Untrustworthy}$
            \ENDIF
        \ENDIF
  \ELSIF {$Direct\ Trust == 1$} 
        \IF {$ (|T| >= |U|) \ \vert \vert \ (|N| >= |T| \ \&\& \ |N| >= |U|)$}
            \STATE Node is Trustworthy
        \ELSE
            \STATE $P_U = \frac{|U|}{Total \ Recommendations + 1}$
            \IF{$P_U > = \theta$}
                \STATE $Node \ is \ \textbf{Untrustworthy}$
            \ELSE
                \STATE $Node \ is \ \textbf{Trustworthy}$
            \ENDIF
        \ENDIF
  \ELSE  
        \IF{$ (|T| > |U|) $}
            \STATE $Node \ is \ \textbf{Trustworthy}$
        \ELSE
            \STATE $Node \ is \ \textbf{Untrustworthy}$
        \ENDIF
  \ENDIF

\end{algorithmic}
\label{alg:final_trust_algorithm}
\end{algorithm}

\begin{figure*}[h]
     \centering
     \begin{subfigure}[b]{0.3\textwidth}
         \centering
         \includegraphics[width=\textwidth]{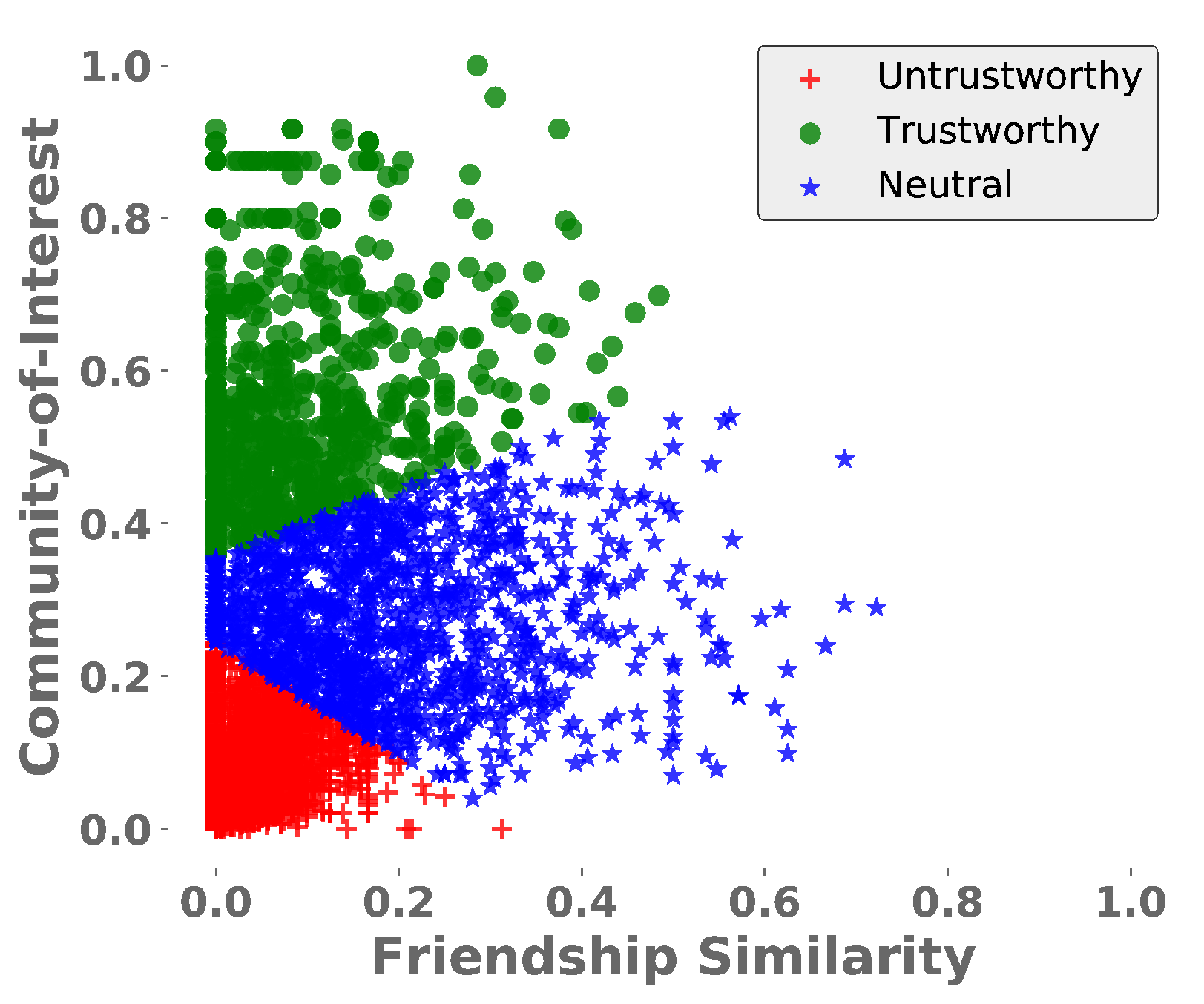}
         \caption{FS and CoI}
         \label{fig:c_coi}
     \end{subfigure}
     \hfill
     \begin{subfigure}[b]{0.3\textwidth}
         \centering
         \includegraphics[width=\textwidth]{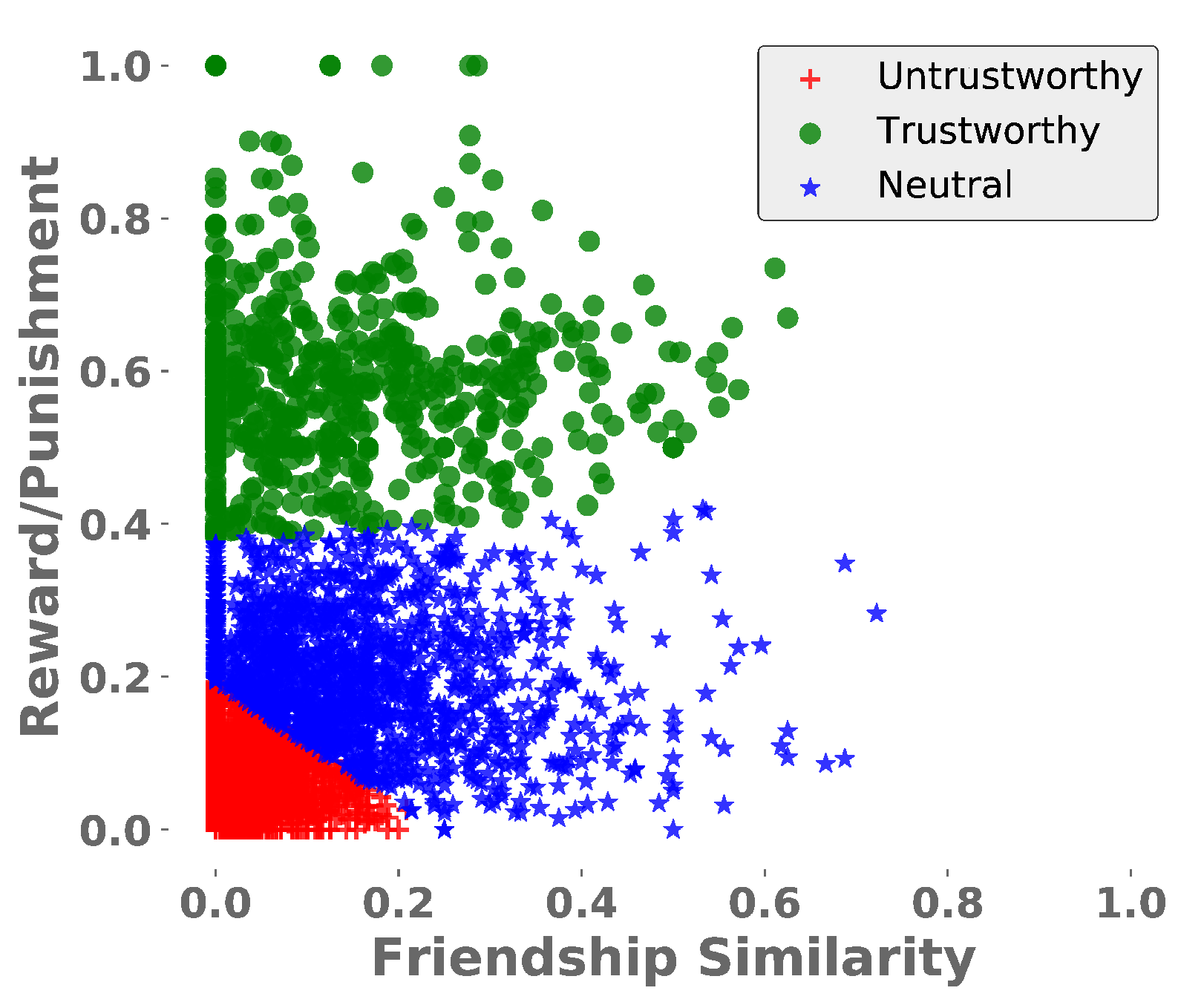}
         \caption{FS and Reward}
         \label{fig:c_rs}
     \end{subfigure}
     \hfill
     \begin{subfigure}[b]{0.3\textwidth}
         \centering
         \includegraphics[width=\textwidth]{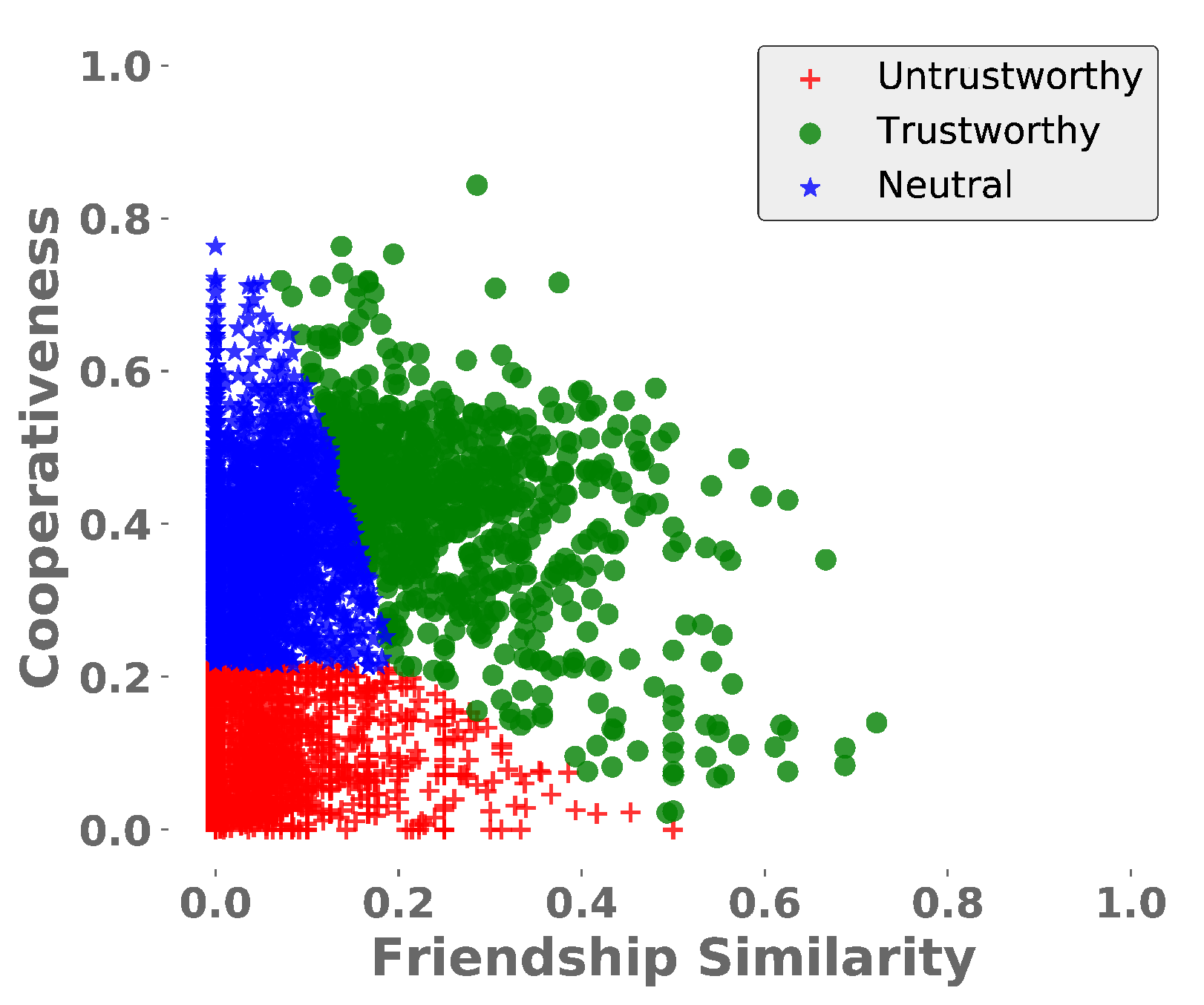}
         \caption{FS and CoP}
         \label{fig:c_cop}
     \end{subfigure}
     \hfill
     \begin{subfigure}[b]{0.3\textwidth}
         \centering
         \includegraphics[width=\textwidth]{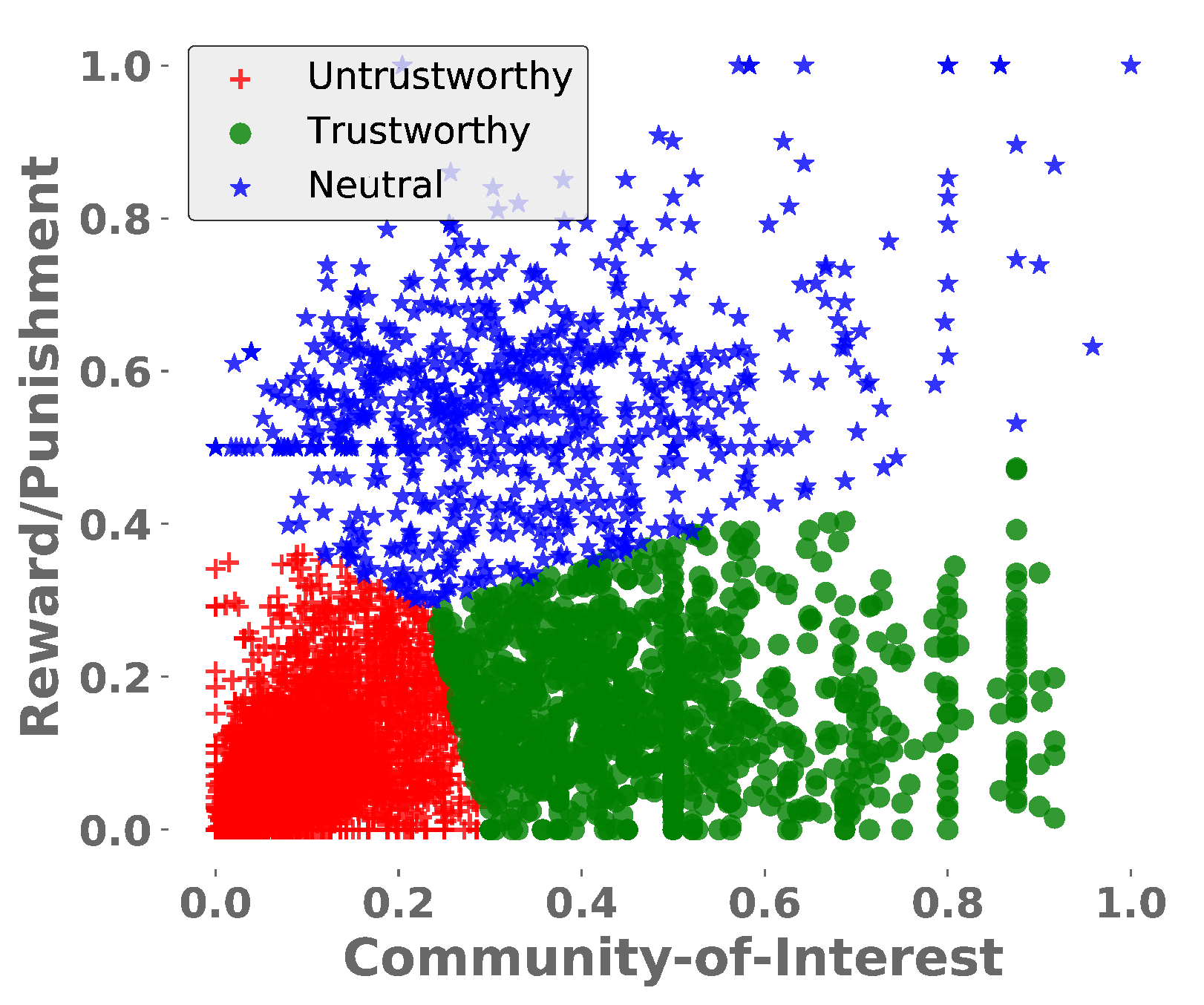}
         \caption{CoI and Reward}
         \label{fig:coi_rs}
     \end{subfigure}
     \hfill
     \begin{subfigure}[b]{0.3\textwidth}
         \centering
         \includegraphics[width=\textwidth]{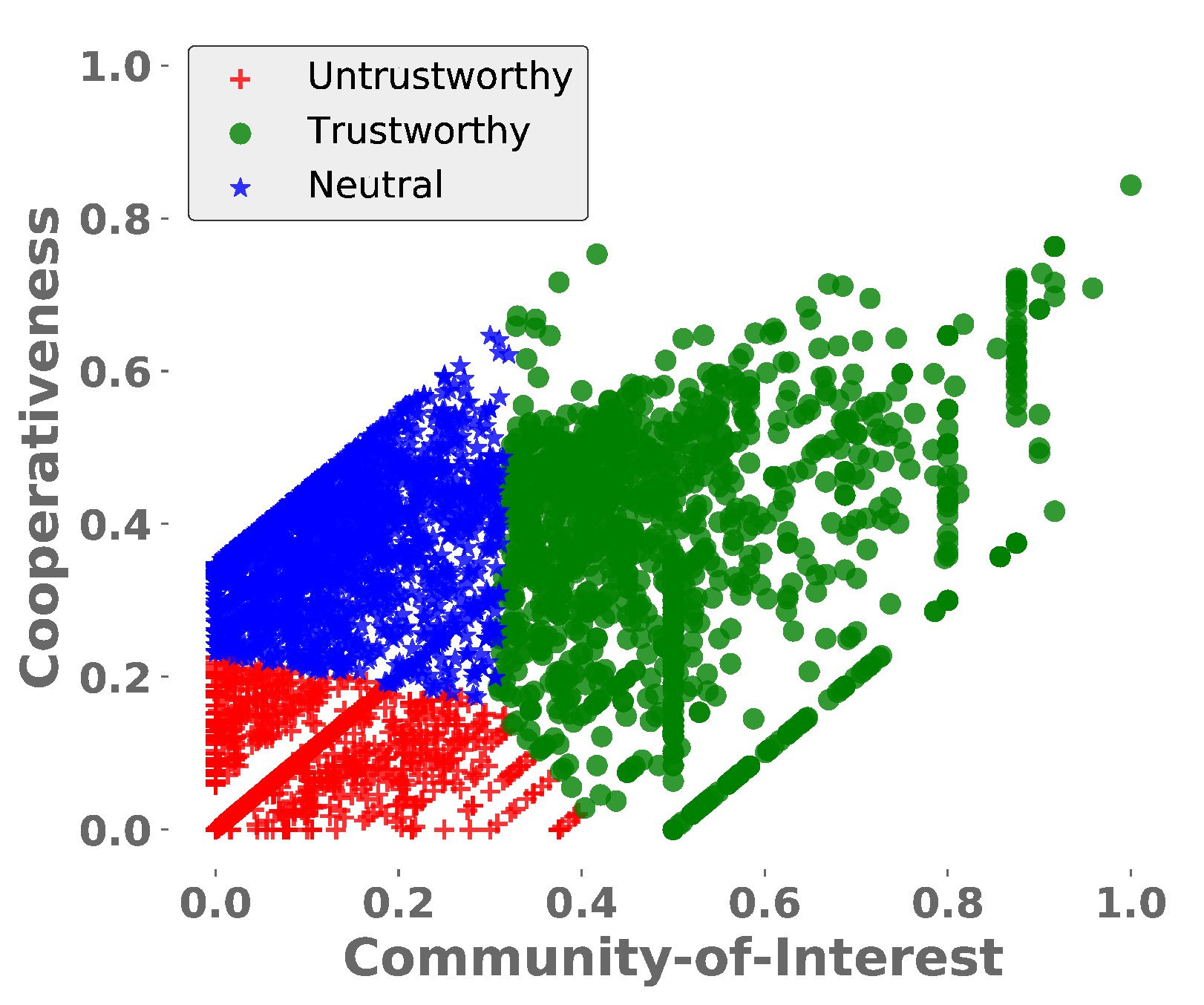}
         \caption{CoI and CoP}
         \label{fig:coi_cop}
     \end{subfigure}
     \hfill
     \begin{subfigure}[b]{0.3\textwidth}
         \centering
         \includegraphics[width=\textwidth]{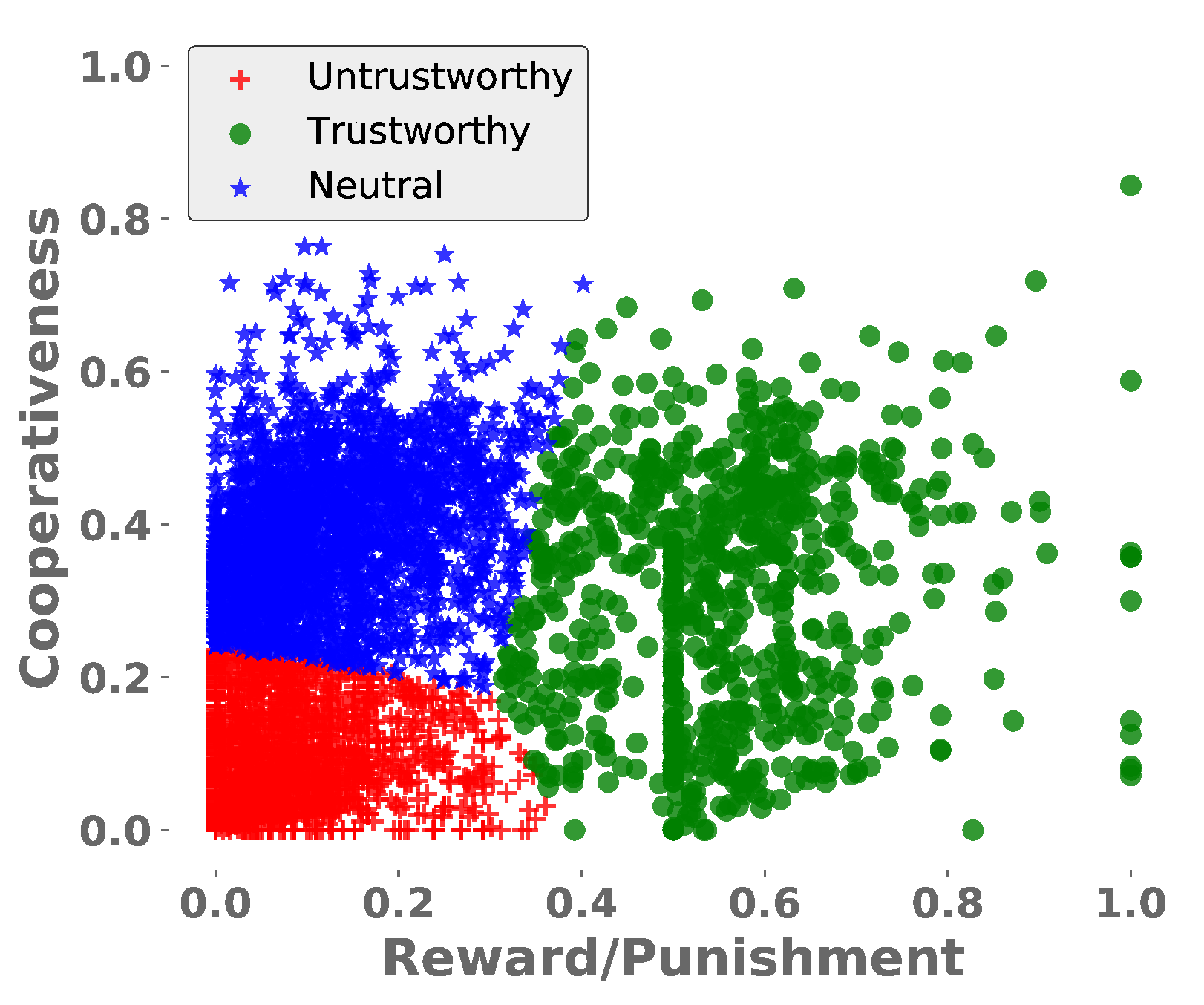}
         \caption{Reward and CoP}
         \label{fig:rs_cop}
     \end{subfigure}
        \caption{Clustering on Different Pairs of Features}
        \label{fig:clustering_pairs}
\end{figure*}

Finally, in order to ascertain a single trust value, we develop an algorithm (i.e., Algorithm 1) so as to aggregate both direct and indirect trust. The said algorithm takes into consideration both of the direct trust and recommendations as an input, and accordingly, provides a \emph{single} trust of a node, i.e., trustworthy or untrustworthy. Our trust score estimation algorithm depends more on direct trust as can be noticed from lines 13-16. If the direct trust is $\textit{0}$ or $untrustworthy$ and more recommendations are untrustworthy ($|U|$) or neutral ($|N|$), the node is marked as untrustworthy. Here, neutral manifests that the node is neither trustworthy nor untrustworthy. Furthermore, if the direct trust is $\textit{0}$ or $untrustworthy$ and the number of trustworthy recommendations are greater than untrustworthy recommendations, i.e., ($|T| > |U|$), then our algorithm does not mark the node as trustworthy immediately. Instead, a percentage of trustworthy recommendations ($P_T$) is computed, and if $P_T$ is greater than threshold ($\theta$), which in our case is $70\% \ or \ 0.7$, then the node is marked as trustworthy (lines 17-22). The value of $\theta$ totally depends on an individual application and the reason for such a high value of $\theta$ in our case is to accord higher authority to the trustor node rather than the recommendations from other nodes in the network to cope with the issue of good mouthing and ballot-stuffing attack. Similarly, this algorithm follows the lines 24-34 provided that the direct trust is $\textit{1}$ or $trustworthy$. At the end, if a direct trust score is $\textit{2}$ or $neutral$, it implies that the trustor node does not possess its own observation for the trustee and the trustworthiness of a node is decided on the basis of recommendations. If $|T| > |U|$, then node is marked as trustworthy, otherwise, it is untrustworthy. 

\begin{figure*}[t]
     \centering
     \begin{subfigure}[b]{0.3\textwidth}
         \centering
         \includegraphics[width=\textwidth]{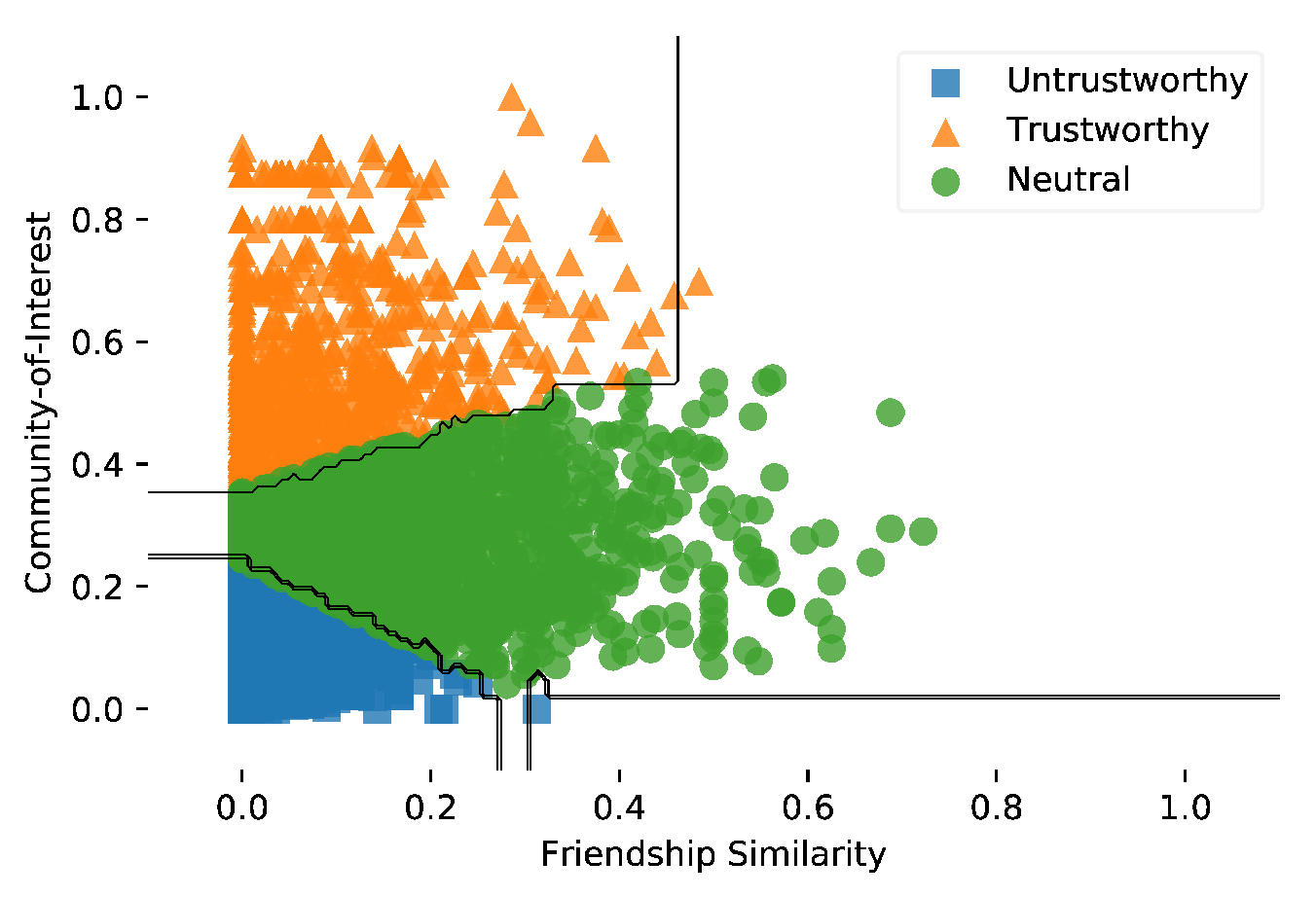}
         \caption{FS and CoI}
         \label{fig:c_coi_d}
     \end{subfigure}
     \hfill
     \begin{subfigure}[b]{0.3\textwidth}
         \centering
         \includegraphics[width=\textwidth]{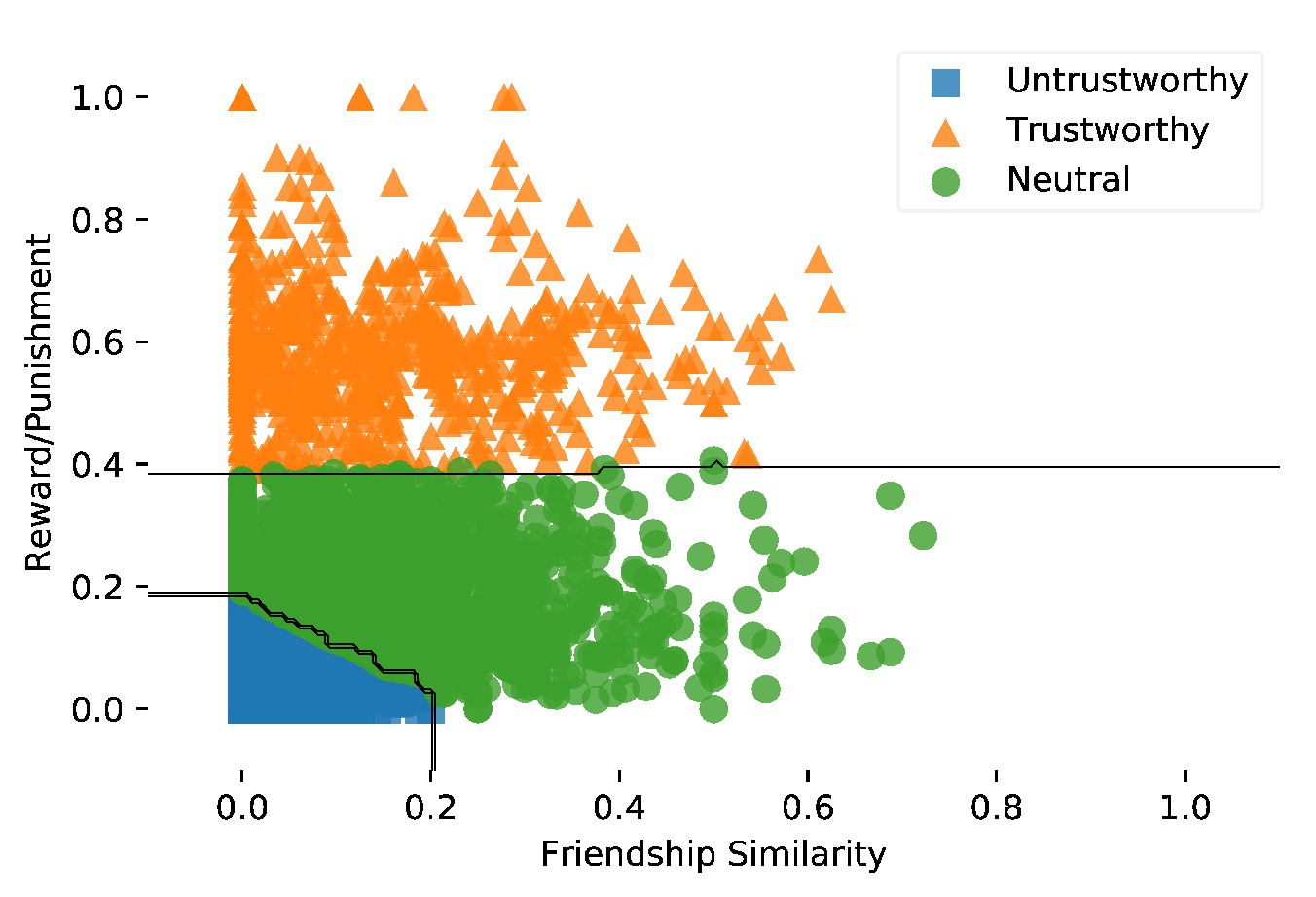}
         \caption{FS and Reward}
         \label{fig:c_rs_d}
     \end{subfigure}
     \hfill
     \begin{subfigure}[b]{0.3\textwidth}
         \centering
         \includegraphics[width=\textwidth]{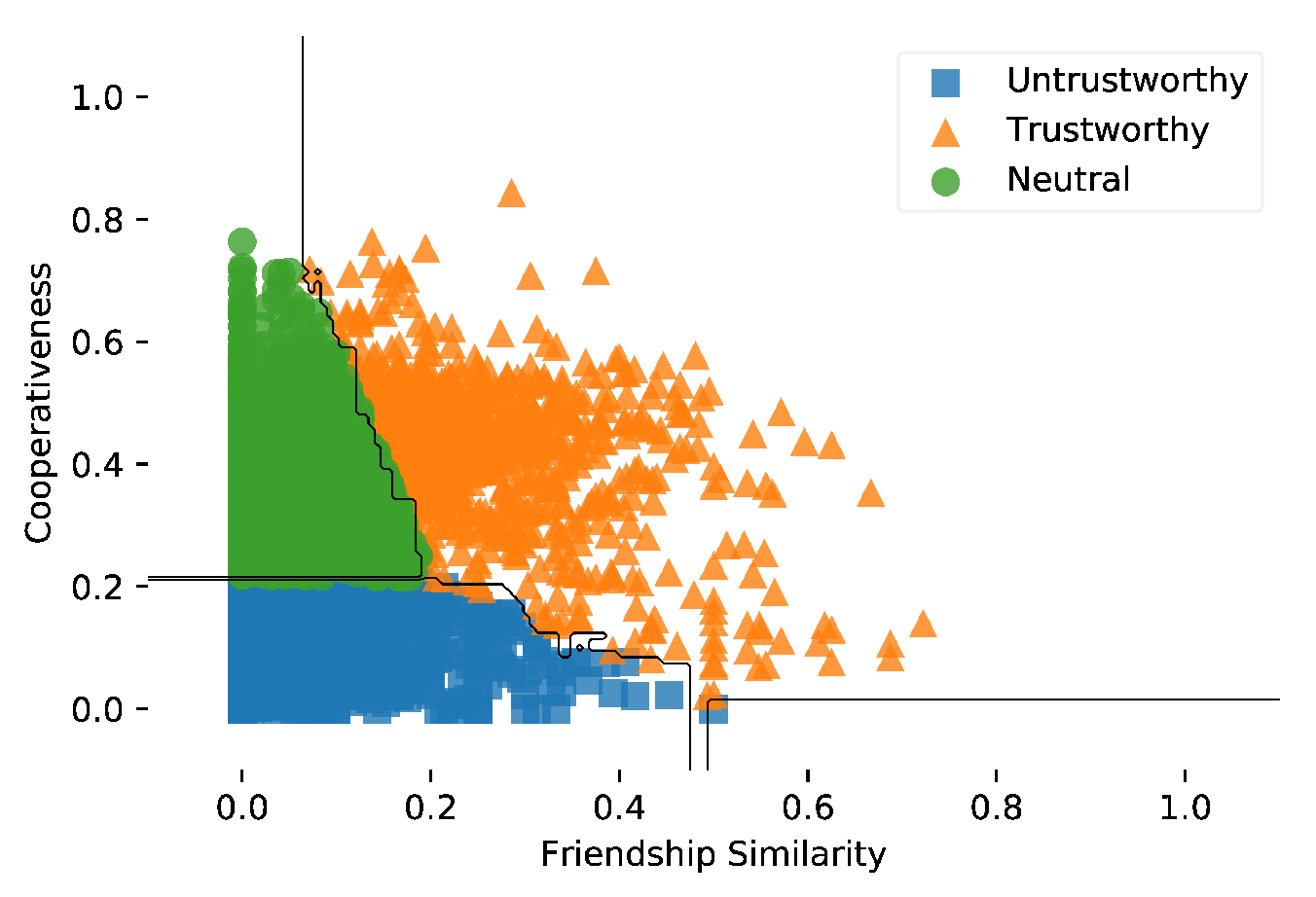}
         \caption{FS and CoP}
         \label{fig:c_cop_d}
     \end{subfigure}
     \hfill
     \begin{subfigure}[b]{0.3\textwidth}
         \centering
         \includegraphics[width=\textwidth]{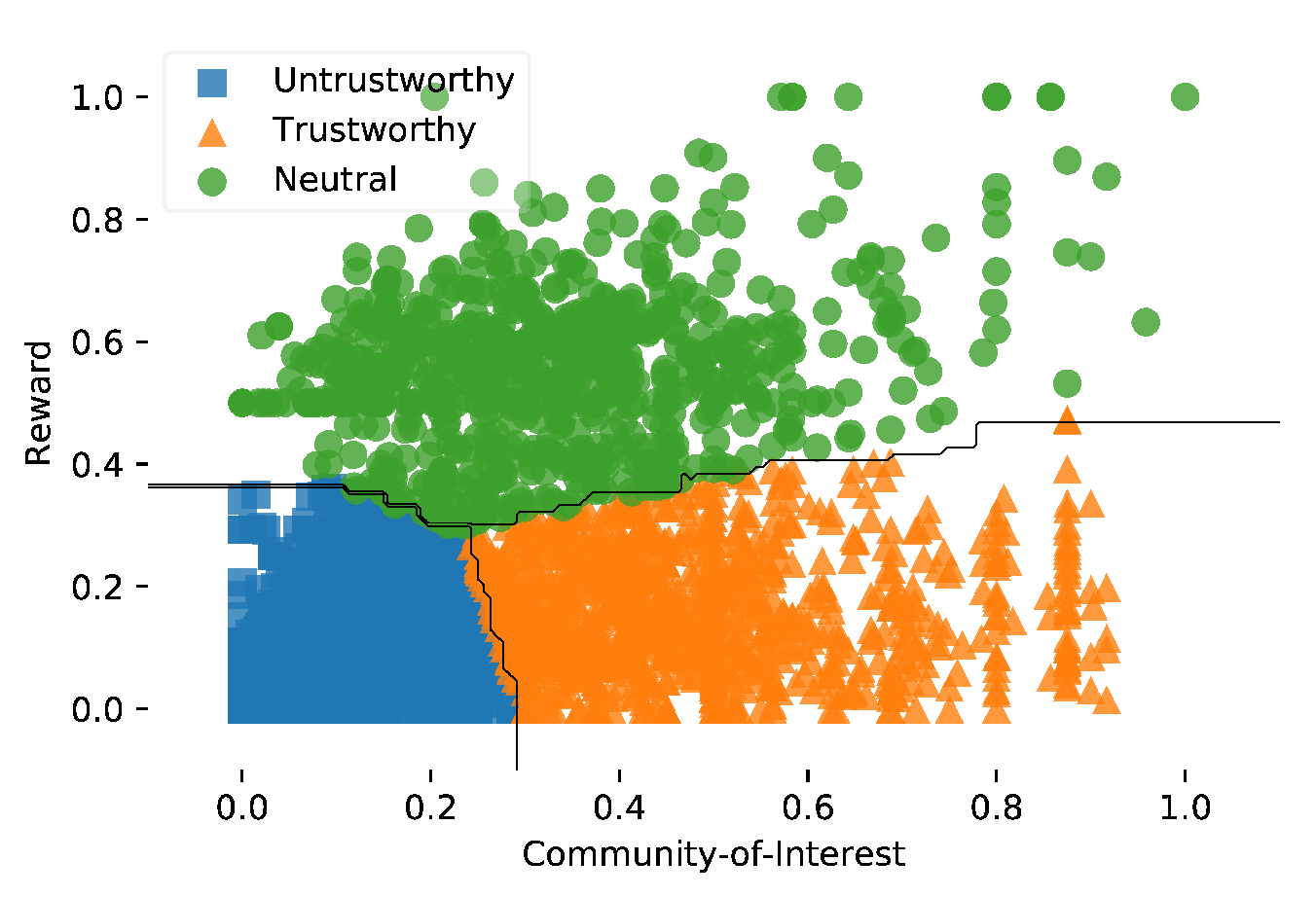}
         \caption{CoI and Reward}
         \label{fig:coi_rs_d}
     \end{subfigure}
     \hfill
     \begin{subfigure}[b]{0.3\textwidth}
         \centering
         \includegraphics[width=\textwidth]{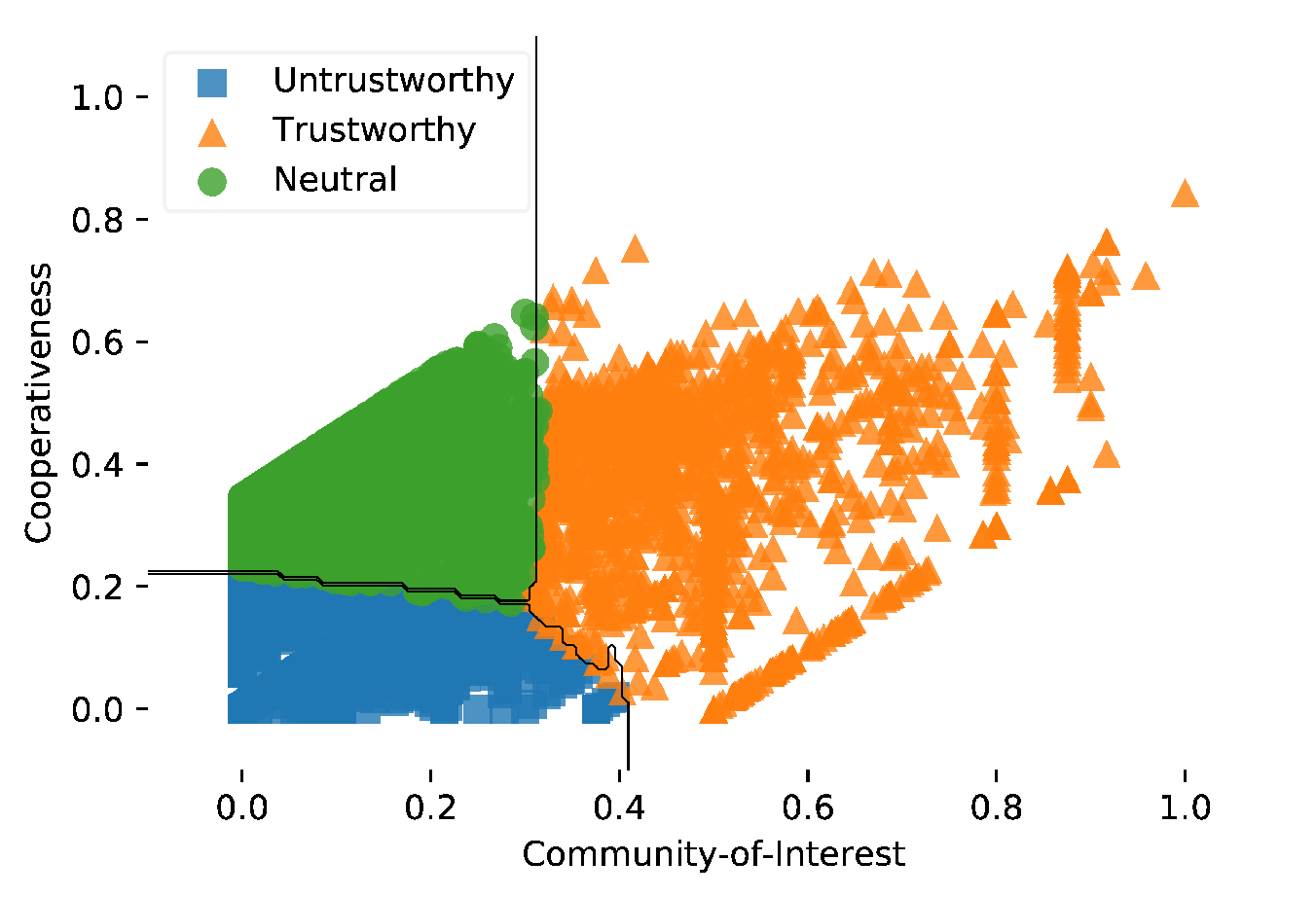}
         \caption{CoI and CoP}
         \label{fig:coi_cop_d}
     \end{subfigure}
     \hfill
     \begin{subfigure}[b]{0.3\textwidth}
         \centering
         \includegraphics[width=\textwidth]{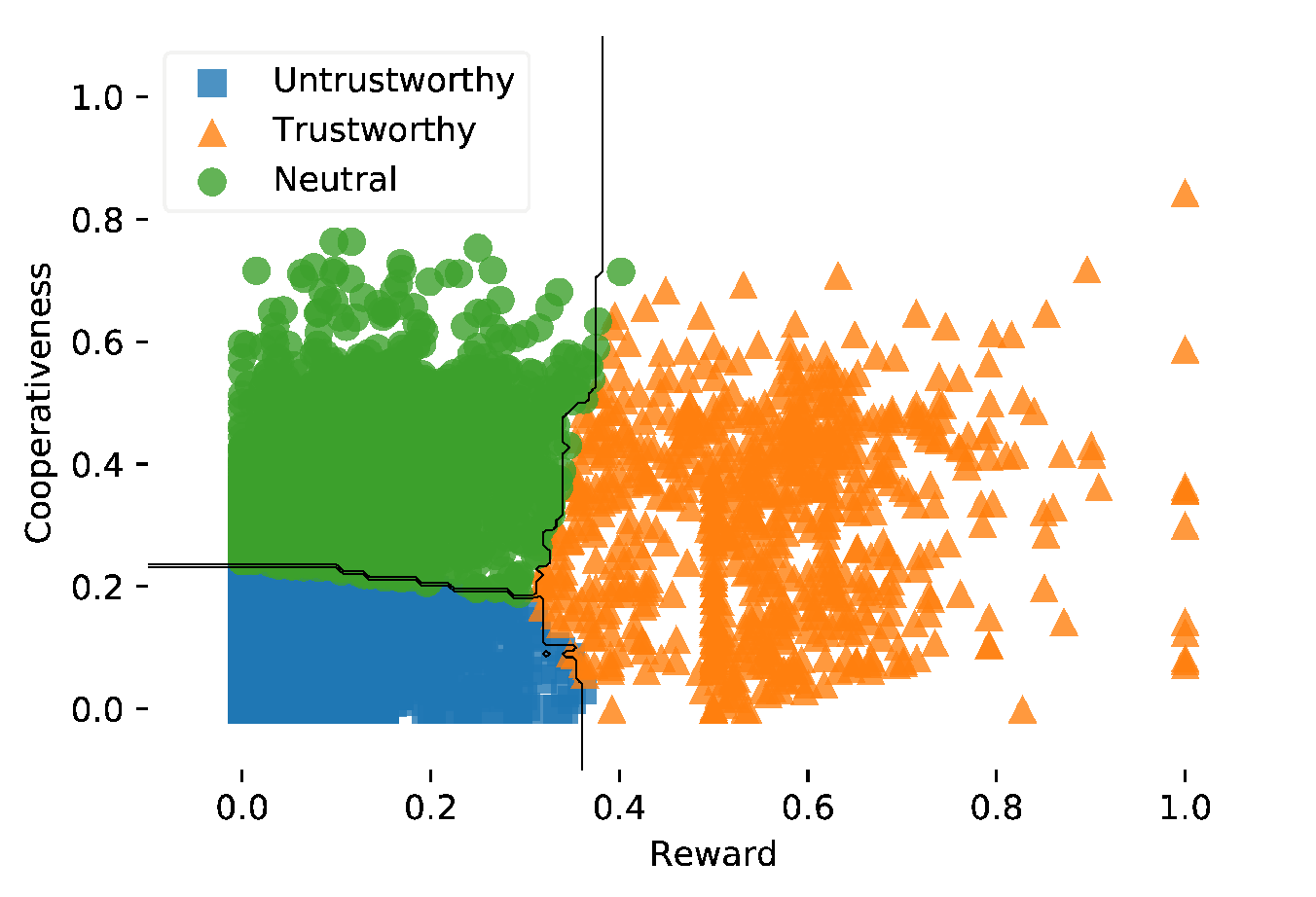}
         \caption{Reward and CoP}
         \label{fig:rs_cop_d}
     \end{subfigure}
        \caption{Decision Boundaries for Different Pairs of Features}
        \label{fig:decision_boundary}
\end{figure*}

\section{Simulation Setup}

To ascertain the aforementioned trust features, i.e., $T_F = \{ FS, CoI, Reward, CoP \}$, to be employed for our ML-based model, we have used \emph{sigcomm-2009}\footnote{\label{note1}https://crawdad.org/thlab/sigcomm2009/20120715/} dataset which comprises the traces that could be mapped in the form of the promising paradigm of SIoT. These traces contain the social information of device/user (i.e., friendships and interested groups information, activities, and messages logs), which have been utilized for computing the trust features mentioned in Section II. This dataset comprises $76$ nodes and $18,226$ interactions for over a period of four days. The trust features are ascertained for each pair of nodes with at least one interaction between them and there are $5,776$ pair of nodes in total.



Subsequent to obtaining of the trust features $T_F$, an unsupervised learning algorithm (i.e., k-means clustering) is employed so as to label the features in different classes \cite{Breiman2001}. The primary reason to use the k-means clustering is due to the unavailability of the labeled training set. The k-means algorithm requires two inputs: initial centroid points ($C$) and the number of clusters ($C_k$). Initially, we have allocated random centroid positions to all the clusters ($C_k$, $k = 1 \ to \ 4$) and executed the algorithm until the convergence. For the proposed model, we needed two clusters, namely trustworthy or untrustworthy, nevertheless, we used the elbow method to acquire the optimized cluster ($C_{opt}$) size with lower k-means cost function, which in our case is, $C_{opt} = 3$. This implies that the k-means algorithm segregates the data in three clusters -- trustworthy, untrustworthy, and neutral, wherein neutral means that the node is neither trustworthy nor untrustworthy and nodes with values near the origin $(0,0)$ are marked as untrustworthy.


In order to train the model subsequent to clustering process, we employed the multi-class random forest classification algorithms \cite{Amatriain2011} to identify the best decision boundary which segregates the trustworthy and untrustworthy interactions. Random forest is fairly suitable for feature engineering, i.e, to identify the most important features amongst all the available features in the dataset, and in our case, it is extremely indispensable to know the more imperious features as well as the weightage of each feature during aggregation of the overall trust. In addition, random forest avoids overfitting problem by constructing multiple decision trees on the same dataset with random selection. For training and testing purposes, $5,776$ samples were utilized which represents the interaction between each pair of nodes. Out of the same, $80\%$ of them were used for training purposes, whereas, $20\%$ of them were used to evaluate the accuracy of the proposed model. 

  
  
  
  
 

\vspace{-2pt}

\section{Results and Discussion}

As deliberated in Section III, k-means clustering has been employed in order to classify the trustworthy and untrustworthy relationships. Nevertheless, it is the elbow method which categorizes the data into three clusters instead of the two, i.e., trustworthy, untrustworthy, and neutral, as portrayed in Fig. \ref{fig:clustering_pairs}. For demonstration purposes, we only took into consideration the pairs of trust features for clustering instead of employing all the features at once since it is not feasible to visualize all of them together. However, we can use the Principle Component Analysis (PCA) technique for dimension reduction, e.g., from four to two in our case, to ascertain the results for visualizing all the features at once \cite{article1}. As can be clearly observed from Fig. \ref{fig:clustering_pairs}a and \ref{fig:clustering_pairs}b, the distribution of trust values by comparing $FS$ with $CoI$ and $FS$ with $Reward$ respectively demonstrate that the region where $CoI >= 0.4$ and $Reward >= 0.4$ are the trustworthy regions, whereas, region where $CoI <=0.2$ and $Reward <= 0.2$ are the untrustworthy regions. However, these figures signify a clear dominance of $CoI$ and $Reward$ on $FS$ as trust value primarily depends on $CoI$ and $Reward$. Furthermore, all other subfigures reflect a mutual contribution to the overall trust score as depicted in Fig. \ref{fig:clustering_pairs}c-\ref{fig:clustering_pairs}f.

\begin{figure}[h]
    \centering
    \includegraphics[width=3.1in, height=2.4in]{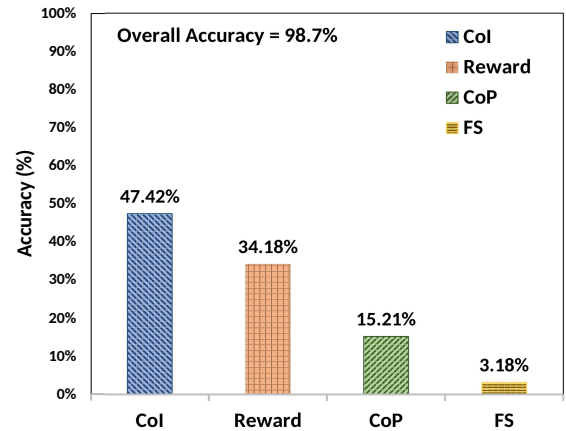}
    \caption{Feature Weightage and Model Accuracy}
    \label{fig:f_imp_acc}
\end{figure}

With successful investigation of the labels, the next step is to train our model to identify whether the futuristic interactions of SIoT nodes are trustworthy or not. After applying a supervised learning algorithm (random forest), Fig. \ref{fig:decision_boundary} portrays the decision boundary to successfully classify the nodes with minimal error. It is quite evident from Fig. \ref{fig:decision_boundary} that our model is fully capable of classifying the futuristic interactions as trustworthy, neutral, or untrustworthy. Fig. \ref{fig:f_imp_acc} manifests the accuracy of the classification algorithm ($98.7\%$) and the significance (i.e., weightage) of each individual feature ascertained after applying our model on the dataset. 
\vspace{-1em}

It can be easily observed that $CoI$ ($47.42\%$) and $Reward$ ($34.18\%$) have more impact on the overall trust score followed by \emph{CoP} ($15.21\%$) and $FS$ ($3.18\%$). The key reason for such a higher weightage of $CoI$ is owing to the fact that the objects belonging to the same group tends to be more trustworthy and interact more frequently. Similarly, the $Reward$ feature shows that more unsuccessful interactions lead to untrustworthiness making it an important factor in calculating the trust score.



As we have identified three categories of nodes, i.e., trustworthy, untrustworthy, and neutral, nevertheless, by and large, we only need to identify a node as either a trustworthy one or an untrustworthy one. To cope with the issue of neutral nodes, we applied our trust score estimation algorithm, i.e., Algorithm 1, on the results obtained from the k-means clustering so as to aggregate the direct trust with the recommendations (indirect trust) and accordingly reduced the size of the clusters to two to be useful for real-world applications. We used a percentage threshold ($\theta$) to combine the trust score, if and when, there is a conflict of interest between the trustor and the other nodes in the network. For instance, if direct trust = $trustworthy$ and the number of untrustworthy recommendations ($|U|$) are more than trustworthy recommendations ($|T|$), then our algorithm investigates the percentage of untrustworthy recommendations ($P_U$). If $P_U>\theta$, then the node is marked as untrustworthy.

\begin{figure}[h]
    \centering
    \includegraphics[width=3.1in, height=2.4in]{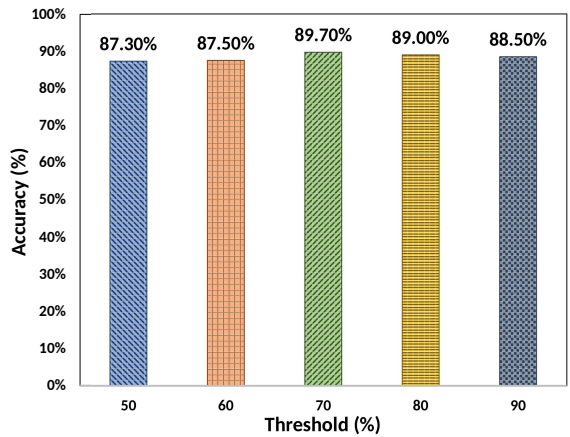}
    \caption{Trust Estimation Accuracy}
    \label{fig:final_acc}
\end{figure}

To ascertain the optimum value of $\theta$, we compared different thresholds, and when $\theta = 70\%$, our algorithm manifests the maximum accuracy of $89.7\%$ as depicted in Fig. \ref{fig:final_acc}. However, the accuracy of our algorithm after combining both direct and indirect trust is bit lower than the accuracy of the model when only the direct trust has been taken into contemplation. This is primarily owing to the fact that there had been no previous history of interactions between the nodes which led to higher trust values at the start. Thus, to avoid such circumstances, recommendations are always employed to ascertain the concrete results which lowers the overall accuracy.  



\section{Conclusion and Future Work}

In this paper, we have proposed a machine learning-based trust aggregation scheme in contrast to the traditional weighted heuristics to ascertain a single trust score for each SIoT node. A trust computational model has been accordingly envisaged to extract key trust features with respect to the SIoT domain. Subsequently, in order to aggregate the trust, the data is labeled by using the k-means clustering for identifying the trustworthy and untrustworthy interactions. A trust prediction scheme has been further proposed for identifying the decision boundaries and to learn the impact of individual features on the aggregated trust score. Our simulation results demonstrate higher accuracy in ascertaining the trustworthy interactions. 

In near future, we intend to incorporate \emph{experience} as a trust attribute for the computation of direct and indirect trust so as to accumulate the previous interaction history of the target nodes coupled with few other social features, i.e., social relationships in terms of co-location and co-work. This could result in more precise determination of trustworthy nodes in a SIoT network.

\bibliographystyle{IEEEtran}
\bibliography{conference_041818}

\end{document}